\documentclass[%
 reprint,
 amsmath,amssymb,
 aps,
]{revtex4-2}

\usepackage{graphicx}
\usepackage{dcolumn}
\usepackage{bm}

\usepackage[table,dvipsnames]{xcolor}

\usepackage{ulem}
\usepackage{pgfplots}
\pgfplotsset{width=10cm,compat=1.9}
\usepgfplotslibrary{fillbetween}
\usepgfplotslibrary{external}
\usetikzlibrary{patterns.meta}
\usepackage{tikz}

\usepackage[bottom]{footmisc}

\begin{document}

\preprint{APS/123-QED}

\title{BPS states in AdS$_3$ Supergravity with Chiral Torsion}


\author{Laura Andrianopoli}
\email[\tt e-mail: ]{laura.andrianopoli@polito.it}
\affiliation{Politecnico di Torino, C.so Duca degli Abruzzi, 24, 10129, Turin - Italy, and\,\\
INFN - Sez. Torino, Via P. Giuria, 1, 10125, Turin, Italy}

\author{Ruggero Noris}
\email[\tt e-mail: ]{noris@fzu.cz}
\affiliation{CEICO, FZU --- Institute of Physics of the Czech Academy of Sciences, 
Na Slovance 1999/2, 182 21 Prague 8, Czech Republic}

\author{Mario Trigiante}
\email[\tt e-mail: ]{mario.trigiante@polito.it}
\affiliation{Politecnico di Torino, C.so Duca degli Abruzzi, 24,  10129, Turin - Italy, and\\
INFN - Sez. Torino, Via P. Giuria, 1, 10125, Turin, Italy}

\author{Jorge Zanelli}
\email[\tt e-mail: ]{jorge.zanelli@uss.cl}
\affiliation{Centro de Estudios Cient\'{\i}ficos (CECs), Av. Arturo Prat 514, Valdivia, Chile,
and \\ Universidad San Sebasti\'an, General Lagos 1163, Valdivia 5110693, Chile}


\date{\today}


\begin{abstract}
In this letter, we construct a supersymmetric model, obtained by deforming $\mathcal N=2$ AdS$_3$ supergravity through a chiral vector component of the torsion. Moreover, we study the existence of BPS states of such theory, by inspecting the presence of Killing spinors on a specific bosonic solution.
\end{abstract}

\maketitle



\section{Introduction}

Pure AdS-Einstein gravity in (2+1) spacetime dimensions has attracted considerable attention from physicists due to its perturbative equivalence to a Chern-Simons theory \cite{Witten:1988hc,Carlip:2005zn,Witten:2007kt}. As gravity here does not describe propagating local degrees of freedom, its richness is encoded in its global properties \cite{Compere:2018aar}. The situation changes when matter sources are included, since they in general spoil the topological properties of the pure Einstein theory and possibly introduce local dynamics. A notable way of including matter in the model, as well as a cosmological constant, is through torsional deformations. Indeed, by properly redefining the torsionful spin connection of the first order formalism, in terms of a torsionless one plus a contorsion, the latter contributes to the energy-momentum tensor in the Einstein equations as a matter source. Torsional deformations in three-dimensional gravity have been extensively studied in the literature, mainly restricting, however, to those choices which do not spoil the local maximal symmetry of the background \cite{Mielke:1991nn,Baekler:1992ab,Mielke:2003xx,Blagojevic:2003vn,Garcia:2003nm,Cacciatori:2005wz,Anninos:2008fx,Alvarez:2011gd,Aviles:2023igk}. This is achieved by only considering the (constant) scalar (i.e. Lorentz singlet) component of the torsion tensor, which, in the absence of a Riemann curvature for the torsionful connection, effectively provides a negative cosmological constant. \\
It has been recently observed in \cite{Andrianopoli:2023dfm}, that, in (2+1) dimensions, one can introduce a vector component of the torsion tensor, defining a non-closed 1-form, together with the scalar one, provided the two are related by a Beltrami-like equation. The latter is an equation for a massive 1-form in three dimensions, which provides the vector component of the torsion with a chiral quality \cite{Townsend:1983xs}. For this reason, we shall also refer to such vector-component as Beltrami-torsion. In the same reference, novel backgrounds featuring these two components of the torsion were constructed and studied. These geometries are not locally ${\rm AdS}_3$ and provide a 1-parameter family of deformations of the BTZ black hole \cite{Banados:1992wn,Banados:1992gq}. \\

Bosonic spacetimes can sometimes be considered as solutions of supersymmetric models, preserving some amount of supersymmetry. In pure supergravities, maximally symmetric spacetimes admit the maximum number of globally defined Killing vectors and preserve the largest number of global supersymmetries. Anti-de Sitter space in (2+1) dimensions, in particular, is a supersymmetric solution of a Chern-Simons model with a supersymmetric extension of the AdS$_3$ isometry group SO$(2,2)$ as its structure group, e.g. the Achucarro-Townsend supergravities with $\mathrm{OSp}(2|p)\times\mathrm{OSp}(2|q)$ \cite{Achucarro:1986uwr}. Supersymmetry of locally ${\rm AdS}_3$ solutions have been extensively studied in the literature, see for example \cite{Coussaert:1993jp,Romans:1991nq,Izquierdo:1994jz}. \\
The goal of the present investigation is threefold:
\begin{itemize}
    \item[1)] to construct a supersymmetric model, based on a torsional deformation of an ${\rm AdS}_3$ supergroup;\vspace{-2mm}
    \item[2)] to show that the geometries considered in \cite{Andrianopoli:2023dfm}, featuring Beltrami-torsion, are solutions to the field equations;\vspace{-2mm}
    \item[3)] to assess the supersymmetry of such backgrounds.
\end{itemize}  
The last objective will be achieved by studying the existence of Killing spinors on the torsionful backgrounds, and comparing the results with the supersymmetric, undeformed ones. Finding supersymmetric solutions in the presence of torsional deformations that behave as matter sources, is somewhat surprising. It can be explained by the fact that the Beltrami torsion defines a new sector, associated with its norm, of which the BPS states are expected to be supersymmetric lowest energy configurations.\\
The letter is organised as follows: in section \ref{sec2}, we explicitly construct a novel supersymmetric model with four supercharges and review the bosonic solution derived in \cite{Andrianopoli:2023dfm}. In section \ref{sec3}, we solve the Killing spinor equation and discuss the global properties of the corresponding solutions, comparing them to the literature \cite{Giribet:2024nwg}. Finally, in section \ref{sec4}, we  comment on the obtained results and  discuss possible future developments.


\section{The geometric setup}\label{sec2}
Let us start by considering the Achucarro-Townsend model \cite{Achucarro:1986uwr}, based on the supergroup OSp$(2|2)\times$SO$(1,2)$ and described by the equations \footnote{We use mostly minus convention for the Lorentz invariant metric $\eta_{ij}$ and define the Levi-Civita symbol as $\epsilon_{012}=\epsilon^{012}=1$.}
\begin{align}\label{ATmodel}
    &R^i\equiv d\omega^i-\frac12\epsilon^i{}_{jk}\omega^j\wedge\omega^k=-i\tau\bar\psi_A\wedge\gamma^i\psi_A\,,\nonumber\\ 
    &\nabla\psi_A\equiv d\psi^A+\frac i2\omega^i\wedge\gamma_i\psi_A+\frac{\tau}{2}\,A\wedge\epsilon_{AB}\psi^B=0\,,\\ \nonumber
    &dA-\epsilon_{AB}\bar\psi_A\wedge\psi_B=0\,,\\\nonumber
    &De^i\equiv de^i-\epsilon^i{}_{jk}\omega^j\wedge e^k=\tau\epsilon^i{}_{jk}\,e^j\wedge e^k+\frac{i}{2}\bar\psi_A\wedge\gamma^i\psi_A\,,
\end{align}
where $i,j, \ldots= 0, 1, 2$ are rigid indices in the vector representation of the local Lorentz group SO$(1, 2)\subset\mathrm{OSp}(2|2)$ and $A,B,\ldots=1,2$ are SO$(2)$ R-symmetry indices. Here $\omega^i=1/2\,\epsilon^i{}_{jk}\omega^{jk}$ is the spin connection of SO$(1,2)\subset\mathrm{OSp}(2|2)$. The connection of the second SO$(1,2)$ factor, say $\omega'^i$, does not appear in the above equations as it is related to the vielbein $e^i$ and the spin connection $\omega^i$ as $\omega'^i=\omega^i+2\tau\, e^i$. The constant $\tau$ defines the AdS$_3$ radius $L$ by $\tau=-\theta/L$, where $\theta=\pm1$. The sign $\theta$, which can be changed by an improper Lorentz transformation, identifies which of the two SO$(1,2)$ factors in the AdS$_3$ isometry group is contained in OSp$(2|2)$, i.e. it determines whether the $(p,q)$ Achucarro-Townsend model is $(2,0)$ or $(0,2)$. The Dirac matrices appearing in \eqref{ATmodel} are written in terms of the Pauli matrices as
\begin{align}
    \gamma^i=\{\sigma_2,i\sigma_1, i\sigma_3\}\,.
\end{align}
In this work, we introduce a torsional deformation of the above superalgebra, effected by changing the last of equations \eqref{ATmodel} into
\begin{align}
De^i&=\tau\epsilon^i{}_{jk}e^je^k+\beta e^i\nonumber \\
&+\left(a\,\epsilon^i{}_{lj}\beta^l+\frac i2\left(1-2ib\,\tau\right)\delta^i_j\right)\bar\psi_A\gamma^j\psi^A\,,\label{deformation}
\end{align}
where $\beta\equiv\beta_i e^i {+\bar\psi_A \lambda^A}$ is a 1-form in superspace, $\lambda_A$ being a Grassmannian field, parametrising the components of $\beta$ in the odd directions. Here and in the following the wedge symbol between differential forms is understood. A more general class of deformations and their consistency conditions are described in Appendix \ref{appA}. \\
The superspace Bianchi identities imply the following relation between the two constants $a$ and $b$
\begin{align}
    i+2\tau\,(b-2 a)=0\,,
\end{align}
together with the following constraints on $\beta_i$ and $\lambda_A$
\begin{align}
    &D_{[p}\beta_{q]}+2\tau\beta^i\epsilon_{ipq}=0\,,\label{ABsuperspace}\\
    &D_{(i}\beta_{j)}=\beta_i\beta_j+2\tau^2\frac{b}{a}\,\eta_{ij}\,,\label{confKilling}\\
    &\lambda_A=\nabla_A\beta_i=0\,,\label{nosusybeta}
\end{align}
where $\nabla_A\beta_i$ is the component of $D\beta_i$ along $\psi^A$, that is $D\beta_i=D_j\beta_ie^j+\bar\psi^A\nabla_A\beta_i$. Relation \eqref{ABsuperspace} was named Beltrami equation in \cite{Andrianopoli:2023dfm}, in analogy with the description of Beltrami flows in fluid dynamics. In fact, it is the equation describing a chiral (anti-)selfdual massive 1-form in three dimensions, see \cite{Townsend:1983xs}, the chirality being related to the sign $\theta$. The spacetime projection of the constraint \eqref{confKilling} on a bosonic background identifies $\beta_i$ as a conformal Killing vector, satisfying the condition: \footnote{A similar condition on the axial component of torsion in four-dimensional conformal gravity was found in \cite{DAuria:2021dth}.}
\begin{align}
\mathring D_i\beta^i=3\left(\beta_i\beta^i+2\tau^2 b/a\right)\,.
\end{align} 
In the supersymmetric case, $\mathring D$ is the covariant derivative with respect to the   ``super-torsionless'' connection $\mathring\omega^i$, that is: 
\begin{align}
  &   \mathring\omega^i=\omega^i +\tau e^i -\epsilon^i{}_{jk}\beta^j e^k\,,   \label{omega0}\\
  & \mathring D e^i=a\,\left(\epsilon^i{}_{lj}\beta^l+2\tau\delta^i_j\right)\bar\psi_A\gamma^j\psi^A\,.
\end{align}
Finally, condition \eqref{nosusybeta} implies that $\beta_i$ does not transform under supersymmetry, that is \begin{align}
    \delta_\varepsilon\beta_i=i_\varepsilon d\beta_i=i_\varepsilon(\psi^A\nabla_A\beta_i)=\varepsilon^A\nabla_A\beta_i=0\,.\label{deltabeta}
\end{align}
If $\beta_i=0$ and $b=0$, the deformation vanishes and we recover the Achucarro-Townsend model in \eqref{ATmodel}. Notice that, if $\beta_i\neq0$, the SO$(2,2)$ AdS$_3$ symmetry is broken. More specifically, the chirality of $\beta_i$ breaks the parity invariance $\theta\to-\theta$. In our framework, the sign $\theta$ accounts for the two inequivalent representations of the Clifford algebra. 
Since the Lorentz-representation of the $\gamma$-matrices is the same as the one of the dreibein (the vector irrep), then the parity transformation $\theta \to -\theta$ could be equivalently accounted for, on a bosonic background, as acting by $e^i \to -e^i$.
In the absence of deformations (that is at $\beta_i=b=0$), both representations define a same bosonic background. In the presence of a non-trivial $\beta_i$, due to the chiral quality of the latter, different signs of $\theta$ identify different bosonic backgrounds.\\
We end this section by writing, for the sake of completeness, besides \eqref{deltabeta}, the supersymmetry transformations of the remaining fields
\begin{align}
    \delta_\varepsilon e^i&=2a\left(\,\epsilon^i{}_{lj}\beta^l+2\tau\delta^i_j\right)\bar\varepsilon_A\gamma^j\psi^A\,,\nonumber\\
    \delta_\varepsilon\omega^i&=-2i\tau\bar\varepsilon_A\gamma^i\psi^A\,,\label{susy}\\
    \delta_\varepsilon\psi_A&=\nabla\varepsilon_A\,.\nonumber
\end{align}
Let us observe that the components $\beta_i$ of the solution derived in \cite{Andrianopoli:2023dfm} in a non-supersymmetric setting, remarkably satisfy both conditions \eqref{ABsuperspace}, \eqref{confKilling}, which are required for a consistent supersymmetric extension of the field equations. This, together with the fact that $\beta_i$ does not transform under supersymmetry, suggests that such bosonic solution, to be reviewed below, may preserve some amount of supersymmetry.

\subsection{The bosonic solution}
A bosonic solution to the above defined supersymmetric model was constructed in \cite{Andrianopoli:2023dfm}. It can be expressed, after proper redefinitions, as: 
\begin{align}
    e^0&=\theta e^0_{\mathrm{AdS}}-\theta\frac{(1-\xi)}{2\xi}\left(dt+\! Ld\phi\right)f\,\frac{2r^2\!-\!r_+^2\!-\!r_-^2}{(r_++r_-)^2}\,,\nonumber\\
    e^1&=\frac{\theta}{\xi}e^1_{\mathrm{AdS}}\,,\nonumber\\
    e^2&=\theta e^2_{\mathrm{AdS}}+\theta\frac{(1-\xi)}{2 \xi}\!\left(dt+ \!Ld\phi\right)\!\frac{2r^2\!-\!r_+^2\!-\!r_-^2}{(r_++r_-)^2}\!\left(\frac{r} {L}\!-\!rN\right)\!,\nonumber\\
    \beta&=\frac{2\epsilon}{r_++r_-}\sqrt{1-\xi^2}\left(f e^0+\left(\frac{r} {L}\!-\!rN\right)e^2\right)\,,\nonumber\\
    A&=0\,,\label{betasol}  
\end{align}
with $\theta$ as defined in the section above,
\begin{align}\nonumber
    e^0_{\mathrm{AdS}}&=f dt, \qquad e^1_{\mathrm{AdS}}=\frac{dr} {f}, \qquad e^2_{\mathrm{AdS}}= r (d\phi+ Ndt)
\end{align}
and 
\begin{align}
   f(r)&=\frac{\sqrt{(r^2-r_+^2)(r^2-r_-^2)}}{Lr}\,,\quad N(r)=-\frac{r_+r_-}{r^2L}\,.
\end{align}
The sign $\epsilon=\pm1$ appearing in \eqref{betasol}, embodies a freedom in the definition of the Beltrami torsion $\beta$. The ranges of the coordinates $(t,r,\phi)$ appearing in the above formulae are as follows: the radial coordinate is positive, $r>0$, whereas $t\in\mathbb R$ and $\phi\in\mathbb R$. The metric in \eqref{betasol} describes a deformation of a locally AdS$_3$ geometry which is parametrised by the non-negative real parameter $\xi$, with $\xi=1$ corresponding to the undeformed background. 
The spacetime admits two Killing vectors, $K_t=\partial_t$ and $K_\phi=\partial_\phi$.\\
The parameters $r_\pm \in \mathbb{C}$ are  integration constants related to the conserved mass and angular momentum of the undeformed geometry by
\begin{align}
    M=\frac{(r^2_++r^2_-)}{L^2}\,,\qquad J=\frac{2r_+r_-}{L}\,.
\end{align}
\\
The quantities in \eqref{betasol}, together with the torsionful spin connection given in Appendix \ref{appB}, satisfy the equations 
\begin{align}
    De^i&=-\frac{\theta} {L}\epsilon^{ijk}e_j\wedge e_k+\beta\wedge e^i\,,\quad R^i=0\,,
\end{align}
together with \eqref{ABsuperspace} and \eqref{confKilling}, with parameters $a=\frac{i}{4\tau\xi^2}$ and $b=\frac{i}{2\tau}\frac{1-\xi^2}{\xi^2}$. These equations then explicitly read
\begin{align}\label{betaeqs}
    \star d\beta=2\frac{\theta}{L}\beta\,, \quad D_{(i}\beta_{j)}=\beta_i\beta_j+4\tau^2(1-\xi^2)\eta_{ij}\,.
\end{align}
When $\xi\neq1$, the invariance $\theta \to -\theta$, enjoyed by the locally AdS$_3$ geometry, is explicitly broken, as the first of \eqref{betaeqs} is not symmetric under such transformation. \\

\begin{figure}[ht]
    \centering

\includegraphics[scale=.9]{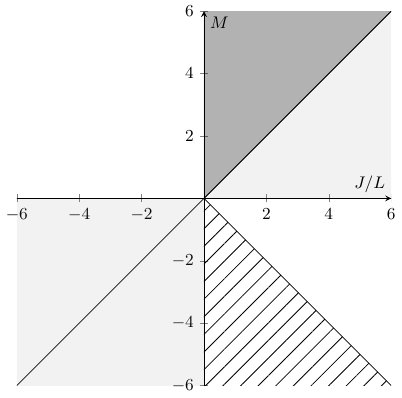}
\caption{We depict here the behavior of the deformed geometry after compactification of $\phi$. White regions correspond to regular geometries, while dark gray regions are black holes. Finally, the light gray region denotes naked singularities, while the white-dashed area corresponds to naked singularities which are not continuously connected with $\xi=1$.}\label{fig1}
\end{figure}

Equations \eqref{betaeqs}, when rewritten in terms of the bosonic restriction of the torsionless spin connection $\mathring\omega_i$ in \eqref{omega0} read
\begin{align}
\mathring{D}_{[i}\beta_{j]}=\frac\theta L\epsilon_{ijk}\beta^k\,,\quad\mathring{D}_{(i}\beta_{j)}=0 \,,  \label{d0beta}
\end{align} 
and identify a proper Killing vector among the admissible conformal ones. 
In fact, one can verify that the vector $\beta^\mu\partial_\mu$, where $\beta^\mu\equiv g^{\mu\nu}\,\beta_\nu$, reads:
\begin{equation}
\beta^\mu\partial_\mu=2\, \theta  \epsilon\,\frac{ \sqrt{1-\xi ^2} }{(r_++r_-)}\,\left(K_t-\frac{1}{L}\,K_\phi\right)\,.\label{betakilling}
\end{equation}
Using \eqref{d0beta}, \eqref{betakilling}, it immediately follows that its norm is constant
\begin{align}
 \label{beta2cost} \partial_i \beta^2&=
2\frac{\theta}L\beta^j\beta^k\epsilon_{ijk}=0\,,
\end{align}
its value being $\beta^2=-4(1-\xi^2)/L^2$. Therefore $\beta^2$ is a characteristic constant of the background, associated with the deformation from a locally AdS$_3$ solution. It can be viewed as an order parameter, selecting a distinct sector of the space of solutions. The Riemann curvature for the torsionless spin connection is
\begin{align}
    \mathring R_0=\frac{1}{L^2}e^1 e^2,\, \mathring R_1=-\frac{1}{L^2}e^0 e^2,\, \mathring R_2=\frac{4\xi^2-3}{L^2} e^0 e^1\,.
\end{align}
The geometry described above may admit the identification $\phi\sim\phi+2\pi$, corresponding to a quotient along the Killing vector $K_\phi$. The quotient is well-defined whenever it does not introduce closed timelike curves. Such causal singularities can be studied by inspecting the reality of the zeroes of the metric component $g_{\phi\phi}(r)$
\begin{align}
    g_{\phi\phi}=-\frac{(1-\xi^2)}{\xi^2(r_++r_-)^2}\,(r^2-\hat r^2_+)(r^2-\hat r^2_-)\,,
\end{align} 
depending on the various parameters. For $\xi\neq1$, the causal singularities are formally located at $r=\hat r_\pm$, where 
\begin{align}
    \hat r_\pm^2=\frac{\left(1-\xi ^2\right) \left(r_+^2+r_-^2\right)-\xi  (r_++r_-)^2\pm \xi\sqrt{\Delta}}{2 \left(1-\xi ^2\right)}\,,
\end{align}
with 
\begin{align}
    \Delta=(r_++r_-)^2 \left(2 \xi ^2 \left(r_+^2+r_-^2\right)-(r_+-r_-)^2\right)\,,
\end{align}
while for $\xi=1$ there is a single zero at $r=0$. When either $\hat r_+$ or $\hat r_-$ are real and covered by a horizon ($r_+>\hat{r}_+$), the spacetime is a black hole, if not it has a naked singularity. When singularities are absent, that is for $\hat r_\pm\not \in \mathbb{R}_{\geq 0}$, the corresponding configurations are regular, albeit not always well defined in the given coordinates \eqref{betasol}.\\
Below we discuss the various cases in which  $M$ and $J$ are real, together with the corresponding behaviour of vielbein, metric and torsion:\vspace{.15cm}\\ 
\textbf{Case A)} \\
$r_\pm\in \mathbb R$, $0<\xi\leq1$, $r_+\geq|r_-|\iff M\geq |J|/L$.\\ 
When $|r_-|<r<r_+$, $e^0$, $e^1$ become imaginary. This can easily be avoided by considering a complex Lorentz rotation
\begin{align}\label{LorentzRot}
    \Lambda=\begin{pmatrix}
        0 & i & 0\\
        i & 0 & 0\\
        0 & 0 & 1
    \end{pmatrix}\,,
\end{align}
which makes the whole solution real and well-defined. We can further distinguish three cases:
\begin{itemize}
    \item[A1)] $r_+> |r_-|\geq0 \iff M>|J|/L$.\\
    For $\xi\neq1$, these  solutions, for strictly positive $r_-$,  which means $J>0$, are black holes;
    \item[A2)] $r_+=|r_-|>0 \iff M=|J|/L$.\\
    Extremal black hole solutions only exist for strictly positive angular momentum. For $\xi\neq1$, $r_+=-r_-$ leads to ill-defined quantities, since it would correspond to a divergent metric and torsion;
    \item[A3)] $r_+=r_-=0\iff M=J/L=0$.\\
    When $\xi\neq1$ the solution is not well-defined in the chosen coordinates.
\end{itemize}
\textbf{Case B)} \\
$r_\pm\in i\mathbb R$, $\xi\geq1$, $\mathrm{Im}(r_+)\geq|\mathrm{Im}(r_-)|\iff M\leq- |J|/L$. \\
The condition $\xi\geq1$ is required by the reality of $\beta$. The above conditions can further be distinguished into:
\begin{itemize}
    \item[B1)] $\mathrm{Im}(r_+)>|\mathrm{Im}(r_-)|\iff M<-|J|/L$.\\
    Causal singularities continuously connected to the solution at $\xi=1$, only exist for $\mathrm{Im}(r_-)\geq0$, that is $J\leq0$. 
    For $J>0$, i.e. for ${\rm Im}(r_-)<0$, causal singularities only occur for $\xi>\xi_0>1$, where $\xi_0$ has the following expression in terms of $r_\pm$ 
    \begin{align}
        \xi_0=\frac{r_+^2+r_-^2}{r_+^2+4 r_+r_-+r_-^2}\,.
    \end{align}
    In this case, the singular solutions are not continuously connected to the locally AdS ones. All these singularities, for $M<0$,  are naked, just as for the undeformed solutions;
    \item[B2)] $\mathrm{Im}(r_+)=|\mathrm{Im}(r_-)|>0\iff M=-|J|/L$.\\
    When $\xi\neq1$, causal singularities do not exist and, as before, $r_+=-r_-$ is not well-defined in the chosen coordinates.
\end{itemize}
\textbf{Case C)} \\
In this case, $r_\pm$ are complex conjugate to each other, up to a sign. We then have:
\begin{itemize}
    \item[C1)] $r_+=r_-^*$, $0<\xi\leq1$, $\iff - J/L<M< J/L,\, J>0$.\\
    When $\xi<1$, causal singularities continuously connected with the solution at $\xi=1$ may exist for $|\mathrm{Re}(r_+)|>|\mathrm{Im}(r_+)|$, which corresponds to $M>0$;
    \item[C2)] $r_+=-r_-^*$, $\xi\geq1$, $\iff J/L<M<-J/L,\, J<0$.\\
    In the $\xi>1$ case, causal singularities exist only for $|\mathrm{Re}(r_+)|<|\mathrm{Im}(r_+)|$, that is for $M<0$.
\end{itemize}
The causal singularities of the metric, due to compactification, are represented in Figure \ref{fig1}.

Let us end this section with a comment.
In the undeformed case, causal singularities always exist, at $r=0$, for any value of $M$ and $J$. By contrast, the above analysis shows that, in the presence of $\beta$, they only exist for specific values of $J$, thus spoiling the symmetry $J\to -J$. More precisely, given a solution, the one with opposite sign of $J$ corresponds to that of the alternative choice of $\theta$ in \eqref{betasol}. \\
Such feature indicates that $\beta$ is capable of regularising the causal structure of the (2+1)-dimensional spacetime, as a consequence of its chiral nature (see the discussion below eq. \eqref{betaeqs}).
This asymmetry was already encountered in \cite{Andrianopoli:2023dfm}, where two inequivalent, deformed solutions with space-like $\beta$, satisfying \eqref{betasol}, were considered, corresponding to positive and negative angular momentum in case A) above. In one case singularities were present, while the other defined a regular solution. 


\section{Killing spinors}\label{sec3}
In this section, we focus on the solution of the Killing spinor equation, in the background shown in \eqref{betasol}, that is on the spinors $\varepsilon_A$ satisfying the requirement $\delta \psi_A=0$ in \eqref{susy}. The equation can be expressed in terms of the Dirac-spinor parameter $\eta\equiv \varepsilon_1+i\varepsilon_2$ as
\begin{align}
d\eta+\frac i2{\omega}_i\gamma^i\eta=0\,.
\end{align}
The strategy for solving the above equation follows the one shown in \cite{Giribet:2024nwg}, which we will briefly outline here. If we define 
\begin{align}
    x^\pm=t\pm L\phi, \qquad \partial_\pm=\frac12\left(\partial_t\pm\frac{\partial_\phi}{L}\right)\,, 
\end{align}
we realise that the spin connection has only components along $dx^+$ {and $dr$}, see Appendix \ref{appB}. Therefore the solution will not depend on $x^-$, that is $\eta=\eta(x^+,r)$.
The equation along $dx^+$ reduces to a single algebraic equation for one of the two components of $\eta$, provided that the spinor is written in terms of independent eigenfunctions of $\partial_+$. Finally, the equation along $dr$ fixes the remaining function.\\
In \cite{Giribet:2024nwg},  the total number of Killing spinors in the undeformed geometry was computed, considering both the two possible inequivalent representations of the three-dimensional Clifford algebra: $\theta\gamma^i$, with $\theta=\pm 1$. We remark that this freedom is the same that we have encoded in our dreibein definition in \eqref{betasol}, with the same meaning for $\theta$. In the deformed case, due to the chirality of $\beta$, only one of the possible values of $\theta$ is allowed, implying, in general, a lower number of Killing spinors with respect to \cite{Giribet:2024nwg}.\\
The solution reads as follows
\begin{widetext}
\begin{align}\nonumber\label{KS}
    \eta(x^+,r)&=e^{-\frac{1}{2L^2}\xi(r_+-r_-)x^+}
    \begin{pmatrix}
    F_+(r;r_-,r_+,\xi)\\
    F_-(r;r_-,r_+,\xi)
    \end{pmatrix}\eta_0+e^{\frac{1}{2L^2}\xi(r_+-r_-)x^+}
    \begin{pmatrix}
    F_+(r;r_+,r_-,\xi)\\
    F_-(r;r_+,r_-,\xi)
    \end{pmatrix}\eta_1\,,\\
    F_\pm(r;x,y,\xi)&=-\sqrt{\frac{(r\pm y) (r\mp x)}{r}} \left(r\pm x-\frac{\xi  (r\mp y)}{1+\epsilon \theta \sqrt{1-\xi ^2}  }\right)\,.
\end{align}
\end{widetext}
The above expression, for $\xi=1$, reproduces the result of \cite{Giribet:2024nwg} (see eq. (3.39) of \cite{Giribet:2024nwg}, with $\theta=-1$). Notice that, for $\xi=1$, the function $F_\pm(r;x,y,1)\propto r^{\tfrac{1}{2}}$ as $r\to \infty$. The Beltrami-torsion, in this Lorentz frame, changes the asymptotic power-law behaviour into $r^{\tfrac{3}{2}}$.\\  
In the following, we shall consider the properties of the above Killing spinors, depending on the parameters $r_\pm$, $\xi$, according to the cases A), B), C) listed in the previous section.\\
The solution \eqref{KS} is globally well-defined, for all considered ranges of coordinates. However, we shall focus on the case in which the coordinate $\phi$ is compact, in the interval $[0,\,2\pi)$. Then, besides ensuring the absence of closed timelike curves, we have to impose that the Killing spinor satisfies periodic/antiperiodic boundary conditions along the compact direction.\vspace{.15cm}\\
\textbf{Case A)}\\
For the Killing spinor to satisfy periodic/antiperiodic boundary conditions along the compact direction, due to the reality of the exponentials, we need to impose the vanishing of the exponents, which requires the extremal limit $r_+=r_->0\iff M=J/L$. This corresponds to case A2) for positive angular momentum. \\
In this case, the background preserves only one supercharge, i.e. it is 1/4 BPS, since the Killing spinor only features one independent  parameter, $\eta_0 +\eta_1$
\footnote{Notice that in the region $|r_-|<r<r_+$, where the dreibein becomes imaginary, we have to multiply the Killing spinor by 
\begin{align}
    S=-\left(
    \begin{array}{cc}
    e^{i\frac{\pi}{8}} & 0 \\
    0 & e^{-i\frac{\pi}{8}} \\
    \end{array}
    \right)\,,
\end{align}
which is the spinorial transformation induced by \eqref{LorentzRot}.
}.\vspace{.15cm}\\
\textbf{Case B)}\\
Compatibility of the Killing spinor solution with the quotient along $K_\phi$ requires imposing a quantisation condition on the exponents, which in this case are purely imaginary: 
\begin{align}
   \xi\Bigl(\mathrm{Im}(r_+)-\mathrm{Im}(r_-)\Bigr)=\xi\sqrt{-\left(M-\frac J L\right)}=kL\,, \label{conditionB}
\end{align}
with $k\in \mathbb Z_{\geq0}.$ The above equation resembles the Dirac quantisation condition on the magnetic monopole charge: in this analogy, the torsion parameter $\xi$ can be seen as the dual to the combination $M-J/L$. Equation \eqref{conditionB} can be rewritten as
\begin{align}\label{MJkeq}
    M=\frac J L-\frac{k^2\,L^2}{\xi^2}\,.
\end{align}
In particular, since $\xi\geq1$, the distance between consecutive lines in the $(M,J)$ plane decreases with the increasing of $\xi$. This  corresponds to cases B1) and B2), the latter only for negative angular momentum. The bosonic solution preserves two supercharges, corresponding to $\eta_0$ and $\eta_1$, i.e. it is 1/2 BPS. \vspace{.15cm}\\
\textbf{Case C)}\\
A Killing spinor solution, satisfying correct boundary conditions, can only be obtained in case C1), where the exponents are complex. This leads to the quantisation rule 
\begin{align}
    2\xi \mathrm{Im}(r_+)=\xi\sqrt{-\left(M-\frac J L\right)}=kL\,,
\end{align}
with $k\in \mathbb Z_{>0}$. The background preserves two supercharges, $\eta_0$ and $\eta_1$, i.e. it is 1/2 BPS.\\

\begin{figure}[ht] 
\centering 

\includegraphics[scale=.9]{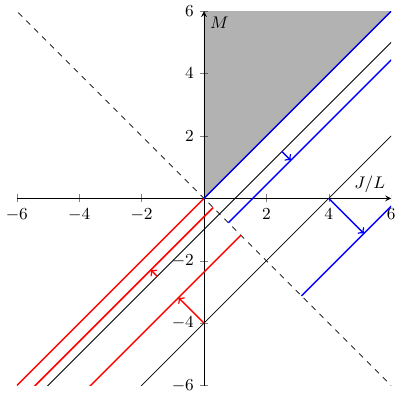}
\caption{We depict here the globally well-defined Killing spinors in the quotient geometries. Diagonal red and blue lines correspond to $\xi=1.3,L=1$ and $\xi=0.8,L=1$, respectively. The arrows highlight the difference between the deformed and $\xi=1$ backgrounds.}\label{fig2} 
\end{figure}

\begin{figure}
\centering
\includegraphics[scale=.9]{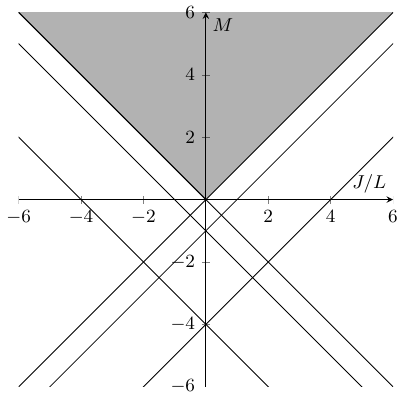}
\caption{Diagonal lines correspond here to the global Killing spinors of the undeformed geometry, which, due to the parity symmetry $\theta\to-\theta$, also admits lines with negative slope.}\label{fig3}   
\end{figure}

We sum up the results of above cases A), B), C) in Figure \ref{fig2}, whereas the underformed configuration is pictured in Figure \ref{fig3}.

\section{Conclusions}\label{sec4}
In this letter, we have constructed a supersymmetric model, by suitably altering the structure of ${\rm OSp}(2|2)\times {\rm SO}(1,2)$ Achucarro-Townsend supergravity. Such deformation is induced by the vector component of the torsion $\beta_i$, which is required to be a conformal Killing vector satisfying the Beltrami equation, besides being a singlet under supersymmetry. The resulting equations are covariant under ${\rm OSp}(2|2)$, while the second ${\rm SO}(1,2)$ symmetry is broken by $\beta_i$, thus spoiling the ${\rm AdS}_3$ isometry group.\\
The backgrounds constructed in \cite{Andrianopoli:2023dfm} are shown to be solution of this supersymmetric model. These geometries are deformations of AdS$_3$, through Killing vectors of the latter. Indeed, the Beltrami torsion happens to define a proper Killing vector $\beta^\mu\partial_\mu$ of both AdS$_3$ and of the deformed background, see eq. \eqref{betakilling}. It comes with no surprise that $\beta_\mu$ satisfies the Beltrami-like equation, since this is a property of all AdS$_3$ Killing vectors. \footnote{This is indeed a general property of the  AdS$_3$ Killing vectors. Writing the isometry group  of this space as ${\rm SO}(2,2)={\rm SL}(2,\mathbb{R})_+ \times {\rm SL}(2,\mathbb{R})_-$ and denoting by $K_{(\pm)x}=K_{(\pm)x}^\mu\,\partial_\mu$ the Killing vectors generating the ${\rm SL}(2,\mathbb{R})_\pm$ factors, respectively, and satisfying the relations $[K_{(\pm)x},\,K_{(\pm)y}]=-\epsilon_{xyz}\,K_{(\pm)}^z$, one can verify that
$$\partial_{[\mu}K_{(\pm)x\vert\,\nu]}=\pm \frac{1}{L}\sqrt{|g|}\,\epsilon_{\mu\nu\rho}\,K_{(\pm)x}^\rho\,.$$ In our solution $\beta^\mu\propto \sqrt{1-\xi^2}\,K_{(\theta)2}^\mu $, where $K_{(\theta)2}$ is a combination of $K_t$ and $K_\phi$ only.} The geometries constructed in this way are shown to preserve, after compactification along $K_\phi$, half the amount of supersymmetry of the underformed ones, due to the chiral nature of the Beltrami torsion. In particular, the only geometries with protected causal singularities which preserve some supersymmetry are extremal black holes with $J>0$ (they are 1/4 BPS). Extremal black holes with negative $J$ of the undeformed theory are regularised by the presence of $\beta_i$ and the resulting geometries do not preserve any amount of supersymmetry.\\

The deformation of more general $(p,q)$ Achucarro-Townsend supergravities, as well as the embedding of the constructed supersymmetric geometries in higher dimensional supergravity have not been considered here and they would deserve a detailed investigation.\\
Moreover, let us observe that the deformation of both AdS$_3$ gravity and supergravity induced by $\beta_i$ was only introduced on-shell, that is at the level of the Maurer-Cartan equations of the superalgebra. In several contexts, which include the AdS$_3$/CFT$_2$ correspondence, an action principle would be desirable. This would allow a concrete study of the two-dimensional quantum theory, dual to the bulk gravity model. This is expected to be a deformation of a conformal field theory, parametrised by the Beltrami torsion $\beta_i$, which is shown to break conformal symmetry of the locally AdS$_3$ background. Moreover, an action principle would allow a consistent definition of finite conserved charges, associated with the existing Killing vectors $K_t$ and $K_\phi$, through renormalisation, as in \cite{Balasubramanian:1999re,Cheng:2005wk,Aoki:2020prb}. This would ultimately allow for a better understanding of the thermodynamics of the susy-preserving extremal black holes mentioned above, as well as of the naked singularities occurring for $M<|J|/L$. This would nicely complement the results discussed in \cite{Briceno:2024ddc}. \\
Finally, as pointed out above, the geometries discussed here feature a Beltrami torsion which is a proper Killing vector, but the supersymmetric model obtained in section \ref{sec2} seems to admit more general deformations, that satisfy the conformal Killing equation. It would then be interesting to look for more general bosonic solutions with this property. 


\begin{acknowledgments}
We acknowledge interesting discussions with Pedro Alvarez, Bianca Letizia Cerchiai, Riccardo D'Auria and Lucrezia Ravera.
R. N. would like to thank Carlo Alberto Cremonini and Joris Raeymaekers for useful discussions and acknowledges GAČR grant EXPRO 20-25775X for financial support. This work was partially supported by ANID/FONDECYT grants 1220862, 1230112 and 1241835.
\end{acknowledgments}

\appendix


\section{A general class of deformations}\label{appA}
Let us consider here, in place of \eqref{deformation}, the following more general extension of the Achucarro-Townsend theory 
\begin{align}\nonumber
    De^i&=\tau\epsilon^i{}_{jk}e^je^k+\frac{i}{2}\bar\psi_A\gamma^i\psi_A+B^{(2)i}\,,
\end{align}
with its integrability condition reading
\begin{align}
    DB^{(2)i}+2\tau\epsilon^i{}_{jk}B^{(2)j}e^k=0\,. \label{generalcondition}
\end{align}
Since $B^{(2)i}$ is a 2-form in superspace, we can postulate the following parametrisation
\begin{align}
    B^{(2)i}=\beta e^i+\bar\psi^AB_A{}^i{}_je^j+ B^i{}_j\bar\psi_A\gamma^j\psi^A\,,
\end{align}
where $B_A{}^i{}_j$ and $ B^i{}_j$ are, in principle, additional fields, contributing to the deformation of the $\mathrm{OSp}(2|2)\times\mathrm{SO}(1,2)$ model. The parametrisation chosen in the main text is instead \textit{rheonomic}, in the sense that it does not introduce additional fields, besides the ``spacetime'' component $\beta_i$. It corresponds to the choice $B_A{}^i{}_j=0$, $ B^i{}_j=a\epsilon^i{}_{lj}\beta^l+b\tau\,\delta^i_j$. \\
The constraint equation \eqref{generalcondition} is a 3-form in superspace, which, once decomposed along the supervielbein basis, yields the following conditions:
\begin{align}
    &D_{[p}\beta_{q]}+2\tau\beta^i\epsilon_{ipq}=0\,,\\
    &\nabla_A\beta_{[l}\delta^i_{m]}-2\beta_jB_A{}^{[i}{}_{[l}\delta^{j]}_{m]}-D_{[l}B_A{}^i{}_{m]}\nonumber\\
    &-\tau B_A{}^i{}_j\epsilon^j{}_{lm}+B_A{}^i{}_{[l}\beta_{m]}-2\tau\epsilon^i{}_{j[l}B_A{}^j{}_{m]}=0\,,\\
    &\mathrm{Tr}(\nabla_AB_B{}^i{}_k)\epsilon^{AB}+(\bar B_{A}{}^{i}{}_jB_B{}^j{}_{k})\epsilon^{AB}=0\,,\\
    &\mathrm{Tr}(\nabla_{(A}B_{B)_0}{}^i{}_k\gamma_l)-(\bar B_{(A}{}^i{}_j\gamma_lB_{B)_0}{}^j{}_k)=0\,,\\
    &D_k  B^i{}_l-i\beta_j\delta^{[i}_l\delta^{j]}_k-2\beta_j B^{[i}{}_l\delta^{j]}_k+\frac12\mathrm{Tr}(\nabla_{A}B^{Ai}{}_k\gamma_l)\nonumber\\
    &-\frac12(\bar B_{A}{}^i{}_j\gamma_lB^{Aj}{}_k)+2\tau\epsilon^i{}_{jk} B^j{}_l=0\,,\\
    &\nabla_A  B^i{}_k=B_A{}^i{}_j\left( B^j{}_k+\frac i2\delta^j_k\right)\,.
\end{align}
Here $\mathrm{Tr}$ denotes a trace along the spinorial indices, whereas $(\,\,)_0$ indicates symmetric-traceless indices. The above derivation relies on the Fierz identity
\begin{align}
\psi^A\bar\psi^B&=-\frac14\mathbf{1}\epsilon^{AB}\bar\psi_C\psi_D\epsilon^{CD}+\frac12\gamma_l\bar\psi^B\gamma^l\psi^A\,,
\end{align}
where $\epsilon_{12}=\epsilon^{12}=1$. Notice that supersymmetry requires the parametrisation of $B^{(2)i}$ to contain at least one of $B_A{}^i{}_j$ and $ B^i{}_j$.\\
These relations gracefully reduce to only three conditions, \eqref{ABsuperspace}, \eqref{confKilling} and \eqref{nosusybeta}, in the case of the \textit{rheonomic} parametrisation \eqref{deformation}.


\section{The Torsionful spin connection }\label{appB}
The spin connection of the bosonic solution \eqref{betasol} reads
\begin{widetext}
\begin{align}\nonumber
    \omega_0&=\omega_{0\mathrm{AdS}}+\frac{f}{L}  \frac{(1-\xi ) \left(2 r^2-r_+^2-r_-^2\right)}{(r_++r_-)^2}\left(dt+Ld\phi\right) +\frac{2\theta\epsilon \sqrt{1-\xi ^2}   (1-L N)}{f {L}\xi  (r_++r_-)}rdr\,,\\ \nonumber
    \omega_1&=\omega_{1\mathrm{AdS}}-f\frac{ 2 {\theta}\epsilon  \sqrt{1-\xi ^2} r  }{L(r_++r_-)}\left(dt+Ld\phi\right)-\frac{2  (1-\xi )}{f L \xi }dr\,,\\
    \omega_2&=\omega_{2\mathrm{AdS}}+\frac{ 2 \theta \epsilon  \sqrt{1-\xi ^2}}{\xi  (r_++r_-)}\,dr+\frac{r}{L}\frac{(1-\xi)(2 r^2-r_+^2-r_-^2)}{(r_++r_-)^2}\left(\frac1L-N\right)\left(dt+Ld\phi\right)\,,\nonumber\\
    \omega_{0\mathrm{AdS}}&=\frac{f}{L}\left(dt+Ld\phi\right), \quad\omega_{1\mathrm{AdS}}=\left(\frac{1}{L}-N\right)\frac{dr}{f}, \quad
    \omega_{2\mathrm{AdS}}=-\frac{r}{L} \left(\frac{1}{L}+N\right)(dt+Ld\phi)\,.
\end{align}    
\end{widetext}

%

\end{document}